\documentclass[aps,prl,twocolumn,floatfix,superscriptaddress,showpacs]{revtex4}

\usepackage{lscape,graphicx}
\usepackage{bbm}
\usepackage{amsmath,amssymb}

\newcommand{\be}{\begin{equation}}
\newcommand{\beq}{\begin{equation}}
\newcommand{\ee}{\end{equation}}
\newcommand{\bea}{\begin{eqnarray}}
\newcommand{\eea}{\end{eqnarray}}
\newcommand{\ba}{\begin{array}}
\newcommand{\ea}{\end{array}}

\begin{document}

\title{Localized versus extended systems in density-functional theory: some lessons
from the Kohn-Sham exact exchange potential}
\author{C. M. Horowitz}
\affiliation{Instituto de Investigaciones Fisicoqu\'{i}micas Te\'oricas y Aplicadas,
(INIFTA), UNLP, CCT La Plata-CONICET, Sucursal 4, Casilla de Correo 16, La Plata, Argentina}
\author{C. R. Proetto}
\altaffiliation[Permanent address: ]{Centro At\'omico Bariloche and Instituto Balseiro,
8400 S. C. de Bariloche, R\'{i}o Negro, Argentina} 
\affiliation{Institut f{\"u}r Theoretische Physik, Freie Universit{\"a}t Berlin,
Arnimallee 14, D-14195 Berlin, Germany}
\affiliation{European Theoretical Spectroscopy Facility (ETSF)}
\author{J. M. Pitarke}
\affiliation{CIC nanoGUNE Consolider, Tolosa Hiribidea 76, E-20018 Donostia, Basque Country, Spain}
\affiliation{Materia Kondenstsatuaren Fisika Saila, Donostia International Physics Center, and Centro F\'{i}sica Materiales CSIC-UPV/EHU,
644 Posta Kutxatila, E-48080 Bilbo, Basque Country, Spain}

\date{\today}

\begin{abstract}
A long-standing puzzle in density-functional theory is the issue of the long-range behavior
of the Kohn-Sham exchange-correlation potential at metal surfaces.
As an important step towards its solution, it is proved here, through a rigurouos asymptotic
analysis and accurate numerical solution of the Optimized-Effective-Potential integral
equation, that the Kohn-Sham exact exchange potential decays as $\ln(z)/z$ far into the vacuum side of an {\it extended} semi-infinite jellium. In contrast to the situation in {\it localized} systems, like atoms, molecules, and slabs, this dominant contribution does not arise from the so-called Slater potential. This exact-exchange result provides a strong constraint on the suitability of approximate correlation-energy functionals.    
\end{abstract}


%71.15.Mb Density functional theory, local density approximation, gradient and other corrections
%31.15.eg Exchange-correlation functionals
%71.10.Ca Electron gas, Fermi gas

\maketitle

\paragraph*{I\lowercase{ntroduction}}
In their seminal density-functional-theory (DFT) investigation of the electronic structure of metal surfaces, Lang and Kohn~\cite{lang} 
pointed out that far outside the surface the Kohn-Sham (KS) exchange-correlation ($xc$) potential $V_{xc}(z)$ 
of DFT should 
behave like the classical image potential
$- \; e^2/4z$, $z$ being the distance from the surface. While from the physical view point this suggestion results attractive and 
reasonable, forty years later its rigorous proof is still an open question. 
Two kind of approaches are possible to address this difficult problem.
One, followed already by some authors~\cite{gunnarsson,almbladh,sham}, is to 
consider exchange and correlation contributions to the KS $xc$ potential {\it together}; 
we note, however, that as the correlation contribution should always be approximated, great 
care must be taken in approximating the corresponding exchange contribution in a {\it compatible} way. 
Within this context, it is not surprising that various asymptotics have been suggested for
the KS exchange-only ($x$-only) potential along this pathway~\cite{gunnarsson,almbladh,sham}. 

A second way to proceed, followed in the present Letter, is to exploit the fact that since the exchange-energy functional is 
known exactly, the corresponding KS exchange potential $V_x(z)$ can also be known exactly, 
by using the Optimized-Effective-Potential (OEP) method of DFT~\cite{grabo}. And knowing 
the exact $V_x(z)$, the analysis of the more elusive KS correlation potential $V_c(z)$ may be 
advanced on firmer grounds than previously. We have succeeded along this second type of approach, 
by proving rigorously that the
asymptotic behavior of the KS exchange potential far into the vacuum side of a semi-infinite 
jellium is of the form $\ln(z)/z$. This analytical result is supported by a fully
self-consistent numerical solution of the OEP integral equation, 
which describes accurately the KS exact exchange potential at the \underline{bulk}, 
\underline{interface}, and \underline{vacuum} regions of our semi-infinite system.

\paragraph{B\lowercase{asic} OEP \lowercase{equations for a metal surface}}
The calculations presented below focus on the semi-infinite (SI) jellium model of a metal surface,
where the discrete character of the positive ions inside the metal is replaced by a uniform
distribution of positive charge (the jellium). The positive jellium density is given by
$n_+(z)=\bar{n} \; \theta(-z)$, which describes a sharp jellium $(z<0)$ - vacuum $(z>0)$ interface at $z=0$.
The model is invariant under translations in the $x,y$ plane (of normalization area $A$), 
so the single-particle KS orbitals of DFT can be rigorously factorized as 
$\varphi_{{\bf k}_{\parallel},k}({\bf r})=e^{i{\bf k}_{\parallel} {\bf \cdot} \boldsymbol{\rho}}\xi_k(z)/\sqrt{A~L}$,
where $\boldsymbol {\rho}$ and ${\bf k}_{\parallel}$ are the in-plane coordinate and wave vector, respectively. $k$ and $L$
refer to the remaining (continuous) quantum number and the normalization length, both along the $z$ direction. 
The effective one-dimensional KS spin-degenerate orbitals $\xi_k(z)$ are 
the solutions of the KS differential equation (atomic units are used throughout)
\begin{equation}
\hat{h}_{\text{KS}}^{k}(z)\xi_k(z):=\left[-\frac{1}{2}\frac{\partial^2}{\partial z^2} +V_{\text{KS}}(z)-\varepsilon_k 
\right]\xi_k(z)=0 \; ,
\label{KSequations}
\end{equation}
where $\varepsilon_k$ are the KS eigenvalues, and $V_{\text{KS}}(z)=V_{\text{H}}(z)+V_{xc}(z)$.
$V_{\text{H}}(z)$ is the Hartree potential, and $V_{xc}(z):=\delta E_{xc}/\delta n(z)$,
with $E_{xc}:=E_{xc}[\{\varepsilon_k\},\{\xi_k\}]$
and $V_{xc}(z)$ being
the $xc$ energy functional and potential, respectively, and $n(z)$ the ground-state electron density.
The OEP integral equation whose solution provides the KS $xc$ potential $V_{xc}(z)$ is
compactly given as~\cite{hpr} 
\begin{equation}
\int_{0}^{k_F} (k_F^2-k^2) \Psi_k^*(z) \xi_k(z) \; dk~+~\text{c. c.} = 0.
\label{OEP}
\end{equation}
 Here,
$\Psi_k(z)$ are the so-called orbital shifts, defined by
\begin{equation}
\Psi_k(z) = \int_{- \infty}^{\infty} \xi_k(z') \Delta V_{xc}^k(z')G_k(z',z)~dz'\; ,
\label{shift1}
\end{equation}
with
\begin{equation}
G_k(z,z') = \frac{1}{\pi} P \int_{0}^{k_F} \frac{\xi_{k'}^*(z)\xi_{k'}(z')}{(\varepsilon_k-\varepsilon_{k'})}~dk'
\label{Green}
\end{equation} 
being the KS Green function, $\Delta V_{xc}^k(z)=V_{xc}(z)-u_{xc}^k(z)$,
and $u_{xc}^k(z)=[4\pi/A(k_F^2-k^2)\xi_k^*(z)]\delta E_{xc} / \delta \xi_k(z)$; $u_{xc}^k(z)$
are usually referred to as orbital-dependent $xc$ potentials. The symbol ``$P$'' in 
Eq.~(\ref{Green}) denotes the ``principal value'', and $k_F$ represents the magnitude of the Fermi
wave vector~\cite{kf}. The {\it exact} $V_{xc}(z)$ entering Eq.~(\ref{OEP}) is obtained as the solution of this
integral equation, which must be solved self-consistently together with Eq.~(\ref{KSequations}).

While for formal discussions the integral form of the OEP equation [Eq.~(\ref{OEP})] is useful,
it is often more convenient to recast it in the
following fully equivalent form, after a well-established sequence of transformations~\cite{grabo,hpr}:
\begin{equation}
V_{xc}(z)=V_{xc}^{\text{KLI}}(z) + V_{xc}^{\text{Shift}}(z) \; ,
\label{OEP*}
\end{equation}
where $V_{xc}^{\text{KLI}}(z)$ represents the so-called Krieger-Li-Iafrate (KLI) contribution~\cite{KLI,note}: 
\begin{equation}
V_{xc}^{\text{KLI}}(z) = \int_{0}^{k_F} \frac{\left| \xi_k(z) \right|^2}{2\pi^2n(z)}  
\left[u_{xc}^{~k}(z) + \overline{\Delta V}_{xc}^{~k}\right]\; \widetilde{dk} \; ,
\label{KLI}
\end{equation}
and
\begin{equation}
V_{xc}^{\text{Shift}}(z) = - \int_0^{k_F}
\frac{\left[ (k_F^2-k^2)\Psi_k(z)\xi_k(z)+\Psi'_k(z)\xi'_k(z)\right]}{2\pi^2n(z)}\widetilde{dk} \; , 
\label{shift}
\end{equation}
with $\widetilde{dk} = (k_F^2-k^2) \; dk$, primes denoting derivatives with respect to the coordinate $z$ 
and the ground-state electron density $n(z)$ being given by the following expression:
\begin{equation}
n(z) = \frac{1}{2\pi^2} \int_0^{k_F} (k_F^2-k^2) \left| \xi_k(z) \right|^2 \; dk \; .
\label{density}
\end{equation}

Now we focus on the exchange contribution $V_x(z)$ to the KS $xc$ 
potential of Eq.~(\ref{OEP*}), as obtained by replacing the orbital-dependent $xc$ potentials $u_{xc}^k(z)$ entering Eqs.~(\ref{KLI})
and (\ref{shift}) by their $x$-only counterparts
$u_x^k(z)$ which are known exactly~\cite{hcpp}. In this case, 
the first term on the rhs of Eq.~(\ref{KLI}) is easily recognized to be twice 
the position-dependent exchange energy per particle $\varepsilon_x(z)$, defined as 
the interaction between a given electron at $z$ and its exact-exchange hole~\cite{hcpp}. 
Noting that $2\varepsilon_x(z)=V_x^{\text{S}}(z)$, $V_x^{\text{S}}(z)$ 
being the so-called Slater potential~\cite{KLI}, we write:
\begin{equation}
V_x(z)=V_x^{\text{S}}(z)+V_x^\Delta(z)+V_x^{\text{Shift}}(z) \; ,
\label{OEP**}
\end{equation}
where $V_x^\Delta(z)$ represents the contribution to the exchange potential $V_x(z)$ from the $x$-only 
counterpart of the second term on the rhs of Eq.~(\ref{KLI}).

\paragraph*{N\lowercase{umerical results}}

In the case of the SI jellium, we have achieved the self-consistent numerical 
solution of the $x$-only counterparts of Eqs.~(\ref{KSequations}) and (\ref{OEP*}). 
The KS equations have been solved by following the procedure explained in Ref.~\cite{hcpp};
the orbital shifts were directly calculated 
from its definition in Eq.~(\ref{shift1}), with the KS Green function computed using the procedure of
Ref.~\cite{1liebsch}. 

The correct asymptotics (at $z\to\infty$) of the Slater potential $V_x^{\text{S}}(z)$ 
have been reported for a SI jellium~\cite{ss1,nastos,qs} and for jellium slabs~\cite{hcpp}, with the result that 
$V_x^{\text{S}}(z)$ decays in both cases as $-\beta / z$, but with a coefficient $\beta$ that in 
the case of a SI jellium is electron-density dependent 
while for jellium slabs $\beta = 1$. 
Hence, here we focus on the remaining contributions: $V_x^{\Delta}(z)$ and $V_x^{\text{Shift}}(z)$.

%\begin{landscape}
\begin{figure}
\includegraphics[width=0.90\columnwidth,angle=-90]{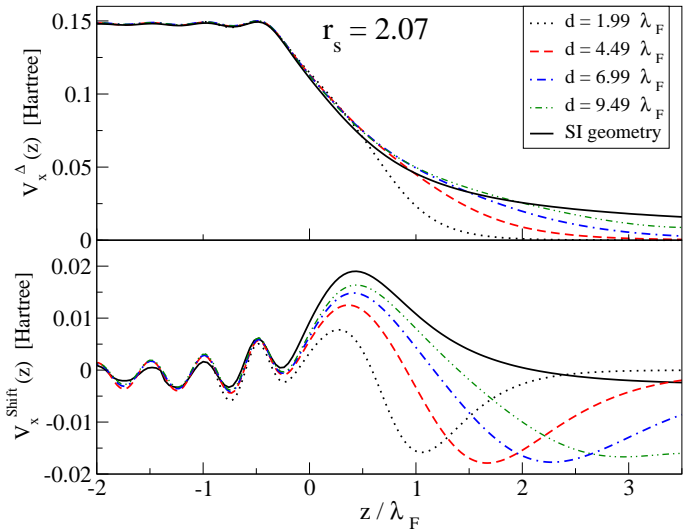}
\caption{(Color online) OEP self-consistent calculations of $V_x^{\Delta}(z)$ (top panel) 
and $V_x^{\text{Shift}}(z)$ (lower panel) for the SI geometry (solid line) and jellium 
slabs of various thicknesses $d$, for $r_s=2.07$. $\lambda_F=2\pi/k_F$ is the Fermi wavelength. 
The jellium-vacuum interface is at $z=0$.}
\label{fig1.ps}
\end{figure}
%\end{landscape}

OEP self-consistent calculations of $V_x^{\Delta}(z)$ and $V_x^{\text{Shift}}(z)$ 
for an electron-density parameter $r_s$ corresponding to the average density of 
valence electrons in Al ($r_s=2.07$) are plotted in Fig.~\ref{fig1.ps}, 
for the SI geometry and for jellium slabs of various thicknesses. 
We note that the bulk limit is correctly reproduced:
$V_x^{\Delta}(z\to-\infty)=k_F/2\pi\simeq 0.148\, {\rm Hartree}$~\cite{gritsenko},
while $V_x^{\text{Shift}}(z\to-\infty)$ 
presents small oscillations around zero, as it should be.
In the case of jellium slabs, the numerical results were 
obtained using the procedure followed in Ref.~\cite{hpr}.

\begin{figure}
\includegraphics[width=0.75\columnwidth,angle=-90]{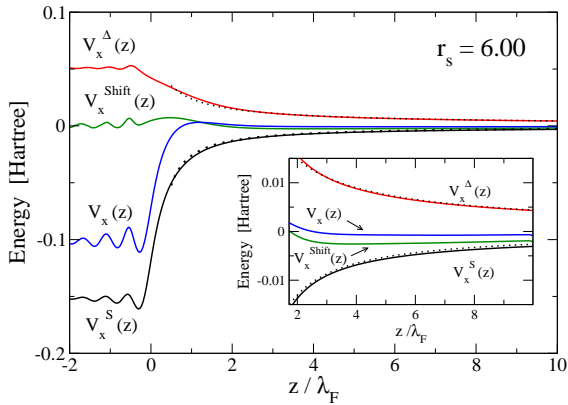}
\caption{(Color online) OEP self-consistent calculations of
$V_x^{\text{S}}(z)$, $V_x^{\Delta}(z)$, and $V_x^{\text{Shift}}(z)$
for the SI geometry and $r_s=6$. The jellium-vacuum interface is at $z=0$. The bulk limits for $V_x^{\text{S}}(z)$, $V_x^{\Delta}(z)$, and $V_x(z)$ are correctly reproduced:
$V_x^{\text{S}}(z\to-\infty)=
-3k_F/2\pi\simeq -0.153\,{\rm Hartree}$~\cite{bardeen},
$V_x^{\Delta}(z\to-\infty)=k_F/2\pi\simeq 0.051\,{\rm Hartree}$, and
$V_x(z \to-\infty)=-k_F/\pi\simeq -0.102\,{\rm Hartree}$. The analytical asymptotes of Eqs.~(\ref{S-expansion}) and (\ref{D-expansion}) are shown by dotted lines. An enlarged view of the far-vacuum region is exhibited in the inset.}
\label{fig2.ps}
\end{figure}

In the vacuum, both $V_x^{\Delta}(z)$ and $V_x^{\text{Shift}}(z)$ decay exponentially for jellium slabs,
as discussed before~\cite{hpr}. In the case of a SI jellium, however, the decay of $V_x^{\Delta}(z)$ and
$V_x^{\text{Shift}}(z)$ turn out to be qualitatively different. This is seen in Fig.~\ref{fig2.ps}, 
where these quantities are plotted for $r_s=6$, together with the Slater potential
$V_x^{\text{S}}(z)$ and the total $V_x(z)$. This figure shows that 
in the case of an extended SI jellium the asymptotics of $V_x(z)$ are dominated by 
$V_x^{\Delta}(z)$, which at large $z$ is positive!

\paragraph*{A\lowercase{nalytical asymptotics}}
In order to determine the actual asymptotic behavior of $V_x(z)$ in the vacuum region of a 
SI jellium, we first appeal to the asymptotic analysis of the KS orbitals $\xi_k(z)$ entering Eq.~(\ref{KSequations}). One finds~\cite{nastos,qs}:
\begin{equation} 
\xi_k(z\to\infty) \rightarrow \xi_{k_F}(z\to\infty) e^{- \alpha z \Delta k}
\label{expansion}
\end{equation}
and
\begin{equation}
n(z\to\infty) \rightarrow \frac{3 \bar{n}}{4(\alpha k_F z)^2}\left| \xi_{k_F}(z\to\infty) \right|^2 \; ,
\label{n-expansion}
\end{equation}
with $\alpha = k_F/\sqrt{2W}$, $W$ being the work function, and $\Delta k = k_F - k$.
This result is perfectly plausible: for $z k_F \gg 1$, the only $k$'s that matter
are those close to $k_F$ and such that $z \Delta k \sim 1$, so the ``window'' for the relevant
$k$'s decreases linearly with distance. As for the electron density, it is interesting 
to note that it decays with an extra power $z^2$ in the denominator that is absent in the case 
of jellium slabs~\cite{hcpp}; this is due to the fact that as $z k_F \gg 1$ the factor $(k_F^2-k^2)$ in the 
integrand of Eq.~(\ref{density}) becomes effectively small, while in the case of jellium slabs 
this factor is always finite and the electron-density decay is purely exponential. 
This anticipates that qualitative differences might be expected between the 
asymptotics of localized (slabs) and extended (SI) systems.

By introducing Eqs.~(\ref{expansion}) and (\ref{n-expansion}) into the expression for the Slater potential [the $x$-only counterpart 
of the first term on the rhs of Eq.~(\ref{KLI})], one finds the known result~\cite{ss1,nastos,qs,hcpp}: 
\begin{equation}
V_x^{\text{S}}(z\to\infty) = - 
\frac{(\pi+2 \alpha \ln \alpha)}{\pi(1+\alpha^2)}\frac{1}{z} \; .
\label{S-expansion}
\end{equation}
Solamatin and Sahni (SS)~\cite{ss1} then derived the asymptotic structure of $V_x(z)$ from an approximate form of
the so-called Sham-Schl\"uter integral equation relating $V_x(z)$ to the non-local Hartree-Fock self-energy and 
concluded that the asymptotics of $V_x(z)$ are embodied by half the Slater 
potential, i.e.: $V_x(z\to\infty)=\frac{1}{2}V_x^{\text{S}}(z\to\infty)$. 
SS supported their result by applying the definition of $V_x(z)$ as the 
functional derivative of the exchange energy (expressed in terms of the Slater potential):
\begin{equation}
V_x({\bf r})=\frac{1}{2}V_x^{\text{S}}({\bf r})+
\frac{1}{2}\int d{\bf r}'\,n({\bf r}')\,\frac{\delta V_x^{\text{S}}({\bf r}')}{\delta n({\bf r})},
\label{ss}
\end{equation}
and then suggesting that the contribution of the second term of Eq.~(\ref{ss}) 
in the vacuum region is zero in the leading order of $1/z$. It is well known, however, 
that the second term of Eq.~(\ref{ss}) contains another term of the form $\frac{1}{2}V_x^{\text{S}}({\bf r})$~\cite{gritsenko}, 
leading, therefore, to an expression for $V_x(z)$ that contains the 
full Slater potential $V_x^{\text{S}}(z)$, and not one half of it, 
as noted by Nastos~\cite{nastos} and 
correctly given in Eq.~(\ref{OEP**}) above. Moreover, here we prove that at large $z$ 
the full $V_x(z)$ of Eq.~(\ref{OEP**}) is not dominated by the Slater 
potential $V_x^{\text{S}}(z)$, but by 
$V_x^\Delta(z)$ instead.

Here we have succeeded, by introducing Eqs.~(\ref{expansion}) and (\ref{n-expansion}) into the $x$-only 
counterpart of the second term on the rhs of Eq.~(\ref{KLI}), to obtain the following neat (positive!) expression for the leading contribution of $V_x^\Delta(z)$ to the long-range exchange potential:   
\begin{eqnarray}
V_x^{\Delta}(z\to\infty) &=& \int_0^{k_F}\frac{\overline{\Delta V}_x^k}
{2\pi^2n(z\to\infty)} 
\left| \xi_k(z\to\infty \right|^2 \; \widetilde{dk} \;\nonumber \\
&=&\frac{1}{2 \pi \alpha z} \left[ \ln(\alpha k_F z) + C \right] \; ,
\label{D-expansion}
\end{eqnarray}
where $C \sim 0.96351$. In passing from the first to the second line we have replaced 
$\Delta V_x^k(z)$, which enters
the calculation of $\overline{\Delta V}_x^k$, by its bulk value. That is, 
$\Delta V_x^k(z) \simeq \Delta V_x^k(z \rightarrow -\infty) = -k_F/ \pi -u_x^k(z \rightarrow - \infty)$.
The explicit (analytic) expression for $u_x^k(z \rightarrow - \infty)$ is obtained 
through a  ${\bf k}_\parallel$ Fourier transform 
of the orbital-dependent exchange potential of a \underline{three-dimensional} electron gas~\cite{bardeen}.

As for $V_x^{\text{Shift}}(z)$, we have
first analyzed the asymptotics of Eq.~(\ref{OEP}) and then applied to it the
operator $\hat{h}_{\text{KS}}^{k_F}(z)$. Solving the resulting equation for $V_x(z\to\infty)$,
one recovers the asymptotic expressions for $V_x^{\text{S}}(z)$ and $V_x^\Delta(z)$ given
by Eqs.~(\ref{S-expansion}) and (\ref{D-expansion}), and one also obtains:
\begin{eqnarray}
&&V_x^{\text{Shift}}(z\to\infty)=\frac{\alpha^2z^2}{k_F\xi_{k_F}(z\to\infty)} 
\int_0^{k_F} \Delta k \; e^{- \alpha z \Delta k} \nonumber \\
&\times& \left[\left(k_F - \frac{\alpha^2 \Delta k}{2}\right) + \alpha\frac{\partial}{\partial z} \right] 
\Psi_k(z\to\infty) \; \widetilde{dk} \; .
\label{shift-potential}
\end{eqnarray} 
At this point, we need $\Psi_k(z\to\infty)$, which we obtain from the
asymptotics of the orbital-shifts differential equation 
$\hat{h}_{\text{KS}}^k(z)\Psi_k(z) = - \left[\Delta V_x^k(z) - \overline{\Delta V}_x^k  \right] \xi_k(z)$.
Noting that at $z k_F \gg 1$ {\it all} contributions in $\Delta V_x^k(z)$ tend to zero, 
the asymptotics of the orbital shifts are found to be given by the following expression:
\begin{equation}
\Psi_k(z\to\infty) \rightarrow \left[f(k) + z g(k) \right] \xi_k(z\to\infty) \; ,
\label{shift-expansion}
\end{equation}
where the first and second terms on the rhs are, respectively, 
the homogeneous and particular solutions of the orbital-shifts differential equation at $z\to\infty$.
Here, $2 g(k) = - \overline{\Delta V}_x^k /(\sqrt{2W}+\alpha \Delta k)$, and the
explicit expression for $f(k)$ is not needed. After introduction of Eq.~(\ref{shift-expansion}) into
Eq.~(\ref{shift-potential}), we find that $V_x^{\text{Shift}}(z\to\infty)$ decays as
$\ln(z)/z^2$. Hence, putting this together with Eqs.~(\ref{S-expansion}) and (\ref{D-expansion}), we conclude that far outside a {\it semi-infinite} jellium the KS exact exchange potential decays as follows
\begin{equation} 
V_x(z\to\infty) =
\frac{\ln(\alpha k_F z)}{2 \pi \alpha z}.
\label{final-expansion}
\end{equation}

Equation~(\ref{final-expansion}) represents the main result of this work.
The asymptotics of Eqs.~(\ref{S-expansion}) and (\ref{D-expansion}) are plotted in Fig.~\ref{fig2.ps} (dotted lines), 
showing that they are in excellent quantitative agreement with our 
fully-self-consistent OEP numerical calculations at $z>\lambda_F$. In retrospective, 
the result of Eq.~(\ref{final-expansion}) looks natural for the SI case. For slabs, 
Eq.~(\ref{OEP**}) yields $V_x(z\to\infty) = -1/z + \overline{\Delta V}_x^m$, $m$ being 
the quantum number corresponding to the highest occupied slab level~\cite{hpr}; the 
first contribution is brought by
the slab Slater potential $V_x^S(z)$, and the second contribution [brought by the 
slab $V_x^{\Delta}(z)$] is a constant which is chosen to be zero~\cite{grabo,hpr}. In 
the SI case, however, while it is still true that the quantity
$\overline{\Delta V}_x^{k_F}$ entering Eq.~(\ref{D-expansion}) is zero, 
the non-negligible contribution from
$\overline{\Delta V}_x^k$ at $k_F-k<1/z$ yields a $V_x^\Delta(z)$ potential that 
decays as $\ln(z)/z$ and dominates the asymptotics of the full $V_x(z)$.        

\paragraph*{C\lowercase{onclusions}}
In summary, we have solved a long-standing problem relative to the long-range behavior of the KS exact exchange 
potential at metal surfaces, as an important step towards the understanding of the actual asymptotic behavior 
of the full KS $xc$ potential. Through a rigurouos asymptotic analysis and accurate numerical solution of 
the OEP integral equation, we have shown that far into the vacuum side of a {\it semi-infinite jellium} 
the KS exact exchange potential decays as $\ln(z)/z$ (positive!). 
This analytical result, which does {\it not} arise from the Slater potential and is supported by a 
fully self-consistent numerical solution of the OEP integral equation, is in contrast to the situation in localized systems, 
like atoms, molecules, and slabs; as in the case of finite systems, for jellium slabs the asymptotics of 
the KS exchange potential arise from the full Slater potential, which decays as $-1/z$~\cite{hpr}. 
Finally, we note that due to the fact that the full KS $xc$ potential of a semi-infinite 
metal should be expected to be absent of the dominant $\ln(z)/z$ exchange asymptotics, 
our exact-exchange result provides a strong constraint on the suitability of approximate correlation-energy functionals.       

\begin{acknowledgments}
C.M.H. acknowledges financial support from CONICET of Argentina.
C.R.P. was supported by the EC's Marie
Curie IIF (MIF1-CT-2006-040222). J.M.P. acknowledges partial support by the University of the Basque Country, the Basque Unibertsitate eta Ikerketa Saila, and the Spanish Ministerio de Educaci\'on y Ciencia (Grants No. FIS2006-01343 and CSD2006-53).
\end{acknowledgments}


\begin{thebibliography}{99}
\bibitem{note}
Mean values are defined as $\overline{O}^i = \int \xi_k^*(z)O^i(z)\xi_k(z) dz$. For simplicity, we choose $\xi_k^*(z)=\xi_k(z)$.
\end{thebibliography}
\end{document}